\documentclass[final]{alt2023}

\title[Algorithms with Predictions for Matching Problems]{Primal-Dual Algorithms with Predictions\\ for Online Bounded Allocation and Ad-Auctions Problems}
\usepackage{times}

\usepackage[utf8]{inputenc}
\usepackage[T5]{fontenc}
\usepackage{caption}
\usepackage{xfrac}
\usepackage{paralist}
\usepackage[capitalise,noabbrev]{cleveref}
\usepackage{algorithm}
\usepackage{setspace}
\usepackage{multicol}

\altauthor{%
 \Name{Kevi Enik\H{o}} \Email{eniko.kevi@univ-grenoble-alpes.fr}\\
 \addr LIG, INRIA, University Grenoble-Alpes, France
 \AND
 \Name{Nguyễn Kim Thắng} \Email{kim-thang.nguyen@univ-grenoble-alpes.fr}\\
 \addr LIG, INRIA, CNRS, Grenoble INP, University Grenoble-Alpes, France
}

\newcommand{\pred}{\texttt{pred}}

\begin{document}

\maketitle

\begin{abstract}%
  Matching problems have been widely studied in the research community, especially Ad-Auctions with many applications ranging from network design to advertising. Following the various advancements in machine learning, one natural question is whether classical algorithms can benefit from machine learning and obtain better-quality solutions. Even a small percentage of performance improvement in matching problems could result in significant gains for the studied use cases. For example, the network throughput or the revenue of Ad-Auctions can increase remarkably. This paper presents algorithms with machine learning predictions for the Online Bounded Allocation and the Online Ad-Auctions problems. We constructed primal-dual algorithms that achieve competitive performance depending on the quality of the predictions. When the predictions are accurate, the algorithms' performance surpasses previous performance bounds, while when the predictions are misleading, the algorithms maintain standard worst-case performance guarantees. We provide supporting experiments on generated data for our theoretical findings.
\end{abstract}


\section{Introduction}

The matching problem is fundamental in combinatorial optimization and operations research with wide applications from student admission to colleges through kidney exchange to Ad-Auctions. Motivated by different markets, for example, advertising and labor markets, online bipartite matching has been intensively studied.
Online bipartite matching revolves around matching a set of items (impressions) to a set of agents (advertisers).
Online items arrive over time, and at the arrival of an item, one needs to make an irrevocable decision on the assignment of the item to an agent.
For unweighted bipartite graphs, \citet{KarpVazirani90:An-optimal-algorithm} gave an elegant algorithm - the \textsc{ranking} algorithm - which always outputs a matching of size at least $(1-1/e)$ times that of the optimal solution \citep{KarpVazirani90:An-optimal-algorithm,BirnbaumMathieu08:On-line-bipartite}.
The ratio of $(1 - 1/e)$ is the best achievable \emph{competitive ratio} in the worst-case paradigm.
Besides, for online matchings in edge-weighted bipartite graphs, no online algorithm is competitive\footnote{except in relaxed models such as disposal-free, etc.} to maximize the total (edge-) weight of the output matching, even though this problem is well-motivated by advertising and AdWords markets (see, for example, \cite{FahrbachHuang20:Edge-weighted-online}).

To circumvent the limitations of the worst-case paradigm, several models have been proposed \citep{Roughgarden19:Beyond-worst-case,Roughgarden20:Beyond-the-Worst-Case}.
The motivation for one of these models is the spectacular advance of machine learning (ML). In particular, the capability of ML methods to predict patterns of future requests could provide valuable information for online algorithms. \citet{LykourisVassilvtiskii18:Competitive-caching} formally introduces a general framework to incorporate ML predictions into algorithms to improve the worst-case performance guarantees. Other researchers followed their work, studying online algorithms with predictions \citep{MitzenmacherVassilvitskii20:Beyond-the-Worst-Case} in a large spectrum of problems, such as
scheduling \citep{LattanziLavastida20:Online-scheduling,Mitzenmacher20:Scheduling-with},
caching (paging) \citep{LykourisVassilvtiskii18:Competitive-caching,Rohatgi20:Near-optimal-bounds,AntoniadisCoester20:Online-metric}
and ski rental \citep{GollapudiPanigrahi19:Online-algorithms,KumarPurohit18:Improving-online,AngelopoulosDurr20:Online-Computation}.
In this paper, we study the design of algorithms with predictions for variants of online edge-weighted matching problems.

\subsection{Model and Problems}
We consider a model (formally stated in \cite{BamasMaggiori20:The-Primal-Dual-method}) in which, whenever a request is released, an algorithm receives some predictions and can use these predictions
to make decisions. The designed algorithms aim to outperform the best-known algorithms when the predictions are correct, but maintain close worst-case guarantees
when the predictions are incorrect. Specifically, we study the following problems in this model.

\subparagraph{Online Bounded Allocation.} In this problem, there are $n$ buyers, each buyer $1 \leq i \leq n$ has a budget $B_{i}$. Items arrive over time; upon the arrival of item $j$,
the price $b_{j}$ of the item as well as the set of buyers $S_{j} \subseteq \{1, 2, \ldots, n\}$ who are interested in purchasing $j$ are revealed. In the standard online setting, one needs to \emph{irrevocably} sell item $j$ to some buyer in $S_{j}$ (or sell it to no one) while respecting the budget constraints of all buyers. In a learning-enhanced model with predictions, each item has a predicted buyer to whom the item should be sold. This additional information may come from a learning procedure based on the data analysis of buyers and items. We consider the learning procedure as a black box and treat the predictions as outputs of an oracle. Concretely, at the arrival of item $j$, the oracle provides a predicted buyer $\pred(j)$ to whom the item $j$ should be sold (conventionally, $\pred(j) = 0$ if the item should not be sold according to the prediction). The objective is to maximize the revenue, which is the total price of sold items.

\subparagraph{Online Ad-Auctions.} This problem is the generalization of Online Bounded Allocation. In this setting, items do not have a fixed price; instead, at the arrival of item $j$, each buyer $i$ proposes a price (or bid) $b_{ij}$ for each item $j$. (Buyers can decline to buy an item by bidding $0$ value.) We assume that the bids are significantly smaller than the buyers' budgets, so $b_{ij} \ll B_{i}~\forall i,j$. Similarly, at the arrival of item $j$, the algorithm receives the prediction $\pred(j)$ and decides to whom to sell the item. The objective is again to maximize the revenue, which is the total price of sold items.

\subparagraph{Learning augmented algorithms.} We aim to design algorithms that incorporate predictions to achieve performance beyond the worst-case bounds. The predictions in our paper can be learned efficiently in practice. Ad-Auctions (or Ad-words) are routinely run a hundred times per day by different search engine companies. At the end of a period (a day, a week, a month), such companies can infer a good matching from their data that better matches the items to the buyers to increase their revenue. In other words, they can leverage large amounts of data to build machine learning models to predict the best buyer for an advertisement slot based on previous transactions. Such predictions can be used in our algorithms.

The learning augmentation depends on the algorithm's confidence in the prediction oracle.  We represent the algorithm's doubt in the prediction by a parameter $\eta \in [0,1]$
and assume that $\eta$ is fixed during the algorithm's execution. Large $\eta$ values mean high doubt, while small $\eta$ values show good confidence.

Let ${\cal A}(I)$ be the objective value of the solution produced by algorithm $\cal A$ on an instance $I$.
Similarly, let $\mathcal{P}(I)$ and ${\cal O}(I)$ be the objective values of the prediction oracle and the optimal solution, respectively.
When the prediction oracle provides an infeasible solution, ${\cal P}(I)=0$.
Given a confidence parameter $\eta \in [0,1]$, we say that an algorithm $\mathcal{A}$ is
 $c(\eta)$-\emph{consistent} and  $r(\eta)$-\emph{robust} if for every instance $I$,
 \begin{align*}
 	\mathcal{A}(I) 	\geq 	\max\{c(\eta) \cdot \mathcal{P}(I),\ r(\eta) \cdot \mathcal{O}(I) \}.
 \end{align*}
Ideally, we would like $c(\eta)$ to tend to $1$ when $\eta$ approaches $0$, meaning that with high confidence, the algorithm performs at least as well as the prediction. Additionally, we would like $r(\eta)$ to tend to the best guarantee as in the standard online setting (without predictions) when $\eta$ approaches $1$.

\subsection{Our Approach and Contributions}
Given an algorithm $A$ that blindly follows the predictions and another algorithm $B$, for example, the best-known algorithm without predictions, a natural question is whether we can derive a new and more efficient algorithm $C$ by taking the linear combination of the two original algorithms. Due to the linear combination, algorithm $C$ can only maintain a consistency of $O(1-\eta)$ for Online Bounded Allocation (or Ad-Auctions), if the robustness is $\Omega(\eta)$ times the worst-case guarantee of algorithm $B$. In this paper, we aim for more substantial and non-trivial guarantees.

We rely on the primal-dual approach to design learning augmented algorithms for both studied problems.
The primal-dual method is an elegant and powerful algorithm design technique \citep{WilliamsonShmoys11:The-design-of-approximation}, especially for online algorithms \citep{BuchbinderNaor09:Online-primal-dual}. To unify previous ad-hoc approaches, \cite{BamasMaggiori20:The-Primal-Dual-method} presented a primal-dual framework to design online algorithms with predictions for covering problems with linear objective functions. However, their approach is not applicable for problems with packing constraints, particularly matching problems and their variants. In this paper, we present learning augmented online algorithms with predictions for the Online Bounded Allocation and the Online Ad-Auctions problems, answering some open questions raised by
\citet{BamasMaggiori20:The-Primal-Dual-method}.

Specifically, in Section \ref{sec:allocation}, we provide an algorithm which is $(1-\eta)$-consistent and \linebreak $\bigl( \frac{e-1}{e} \cdot \frac{1}{1 + (1-\eta)(1 - e^{\eta-1})} \bigr)$-robust
for the Online Bounded Allocation problem. In Section \ref{sec:ad-auction}, we give a $(1-\eta)$-consistent and $(1-e^{-\eta})$-robust algorithm for the Online Ad-Auctions problem. Similarly to \cite{BamasMaggiori20:The-Primal-Dual-method}, we provide algorithms that produce fractional solutions for the studied problems. When the algorithms have high confidence in the prediction (meaning that $\eta$ is closed to 0), the algorithms achieve a similar objective value as the prediction; and when the confidence is low ($\eta$ is close to 1), the algorithms guarantee the best worst-case bound, $(e-1)/e$. This robustness guarantee holds by our algorithms even if the predictions give an infeasible solution (realized during the execution).

\subsection{Related work}

The domain of algorithms with predictions \citep{MitzenmacherVassilvitskii20:Beyond-the-Worst-Case} - or learning augmented algorithms - has recently emerged and rapidly grown at the intersection of (discrete) algorithm design and machine learning (ML). Its main concept is to incorporate learning predictions and ML techniques in the algorithm design to achieve performance guarantees beyond the worst-case analysis and provide tailored solutions to different problems. Recent studies showed interesting results over a large spectrum of problems, such as
scheduling \citep{LattanziLavastida20:Online-scheduling,Mitzenmacher20:Scheduling-with},
caching (paging) \citep{LykourisVassilvtiskii18:Competitive-caching,Rohatgi20:Near-optimal-bounds,AntoniadisCoester20:Online-metric},
ski rental \citep{GollapudiPanigrahi19:Online-algorithms,KumarPurohit18:Improving-online,AngelopoulosDurr20:Online-Computation},
counting sketches \citep{HsuIndyk19:Learning-Based-Frequency} and
bloom filters \citep{KraskaBeutel18:The-case-for-learned,Mitzenmacher18:A-model-for-learned}. \citet{BamasMaggiori20:The-Primal-Dual-method} recently proposed a primal-dual approach to design online algorithms with predictions for linear problems with covering constraints. They raised an open question to extend their approach to packing constraints, which we answer in this paper.

The online matching and Ad-Auctions problems have been widely studied (see \citet{Mehta13:Online-Matching} and references therein).
\citet{MehtaSaberi07:Adwords-and-generalized} introduced the Online Ad-Auctions problem and gave an optimal $(1-1/e)$ competitive ratio when
$R_{\max} = \max_{i,j} \{b_{ij}/B_{i}$\} is small. \citet{BuchbinderJain07:Online-primal-dual} simplified their work by a primal-dual analysis. In the same paper,
the authors gave an algorithm with a refined competitive ratio for the Online Bounded Allocation problem assuming an upper bound on the maximum degree of the vertices in the matching.
\citet{AggarwalGoel11:Online-vertex-weighted} studied another particular case of Ad-Auctions, in which for each $i$ the bids ($b_{ij}$) are the same for every $j$.
They obtained the optimal $(1-1/e)$ competitive ratio with the generalization of the \textsc{ranking} algorithm \citep{KarpVazirani90:An-optimal-algorithm}.
In the Ad-Auctions problem (without any assumptions), the existence of an $(1-1/e)$-competitive algorithm has been conjectured, but remained an open problem.
\citet{HuangZhang20:Adwords-in-a-Panorama} recently presented a $0.5016$-competitive algorithm for this problem.

Motivated by internet advertising applications, several works considered the Ad-Auctions problem in various settings where forecasts or predictions are available or learnable. \citet{EsfandiariKorula18:Allocation-with} proposed a model in which the input is stochastic, and the model gets a forecast for future items. Intuitively, we can measure the forecast accuracy by the optimal solution's fraction one can obtain from the stochastic input. They provide algorithms with provable bounds in this setting. \citet{SchildVee19:Semi-Online-Bipartite} introduced a semi-online model in which the unknown future has a predicted and an adversarial part. They gave algorithms with competitive ratios depending on the fraction of the adversarial part in the input. Closely related to our work is the model by \citet{MahdianNazerzadeh12:Online-Optimization} in which, given two algorithms, one needs to design a (new) algorithm that is robust to both given algorithms. They derived an algorithm for the Ad-Auctions problem that achieves a fraction of the maximum revenue of the given algorithms. The main difference compared to our model is that their algorithm is \emph{not} robust if one of the given algorithms provides infeasible solutions (which could happen with predictions), whereas our algorithm is.

\section{An Algorithm with Predictions for Online Bounded Allocation}   \label{sec:allocation}

Recall that in this problem there are $n$ buyers and each buyer $1 \leq i \leq n$ has a budget $B_{i}$. Upon the arrival of item $j$,
the algorithm discovers the item's fixed price $b_j$ and the subset of buyers $S_j$ interested in purchasing the item. Additionally, the algorithm gets a predicted buyer $\pred(j)$ to whom item $j$ should be sold ($\pred(j) = 0$ if the item should not be sold according to the prediction).
We are interested in fractional solutions as in \citet{BamasMaggiori20:The-Primal-Dual-method}, so we consider the items to be splittable. Before the arrival of the next item, the algorithm needs to make an \emph{irrevocable} decision
and sell the current item in some fractions to some buyers.
The objective is to gain maximum revenue from selling items to buyers.

\subparagraph{Formulation.}
Let $x_{ij}$ be the fraction of item $j$ sold to buyer $i$.
The problem can be cast to the following primal linear program in  \cref{fig:formulation1} (in which we also show its dual program).

\begin{figure}[h]
\centering
\begin{minipage}[t]{0.4\textwidth}
Primal:
\begin{align*}
\max\ \sum_{j=1}^{m} \sum_{i \in S_j}\ & b_j\ x_{ij} \\
\sum_{j: i \in S_j}\ b_j\ x_{ij} &\leq B_i && \forall\ i & (y_i)\\
\sum_{i \in S_j}\ x_{ij} &\leq 1 && \forall\ j & (z_j)\\
x_{ij} &\geq 0 && \forall\ i,\ j & \\
\end{align*}
\end{minipage}
\quad
\begin{minipage}[t]{0.48\textwidth}
\quad Dual:
\begin{align*}
& & \min\ \sum_{i=1}^{n}\ B_i\ y_i &+ \sum_{j=1}^{m}\ z_j  \\
& & b_j\ y_i + z_j &\ge b_j && \forall\ j,\ i \in S_j & (x_{ij})\\
& & y _i &\ge 0 && \forall\ i & \\
& & z_j &\ge 0 && \forall\ j & \\
\end{align*}
\end{minipage}
\caption{Formulation of the Online Bounded Allocation problem}
\label{fig:formulation1}
\end{figure}



\subparagraph{Algorithm.} \citet{BuchbinderJain07:Online-primal-dual} proposes an algorithm for the Online Bounded Allocation problem without predictions. They define \emph{buyer levels} based on the fraction of the buyers' spent budget. Their algorithm splits each arriving item $j$
equally among its interested buyers $i\in S_{j}$ who are on the lowest level. Intuitively, this algorithm corresponds to water-filling the buyer levels.

We propose an algorithm that allocates items using the water-filling strategy of \cite{BuchbinderJain07:Online-primal-dual} and also subtly incorporates predictions.
Recall that the parameter $\eta$ represents the confidence in the predictions. When $\eta$ is close to $1$, our algorithm resembles the water-filling algorithm, while with $\eta$ close to $0$, the prediction has a stronger impact on the algorithm's decision. To do water-filling, our algorithm uses buyer levels as well. At any moment during the execution, a buyer $i$'s level is $\ell$, if the fraction of buyer $i$'s spent budget is within the range of $\left[\frac{\ell}{d} B_{i}, \frac{\ell+1}{d} B_{i}\right)$.

We formally describe the algorithm as follows. Upon the arrival of a new item $j$, let $i^{*}$ be the predicted buyer ($\pred(j)$) suggested by the oracle. Do the following.

\hfill

\begin{spacing}{0.1}
\noindent\rule{\textwidth}{0.5pt}

\noindent \textbf{Algorithm 1} Learning Augmented Algorithm for the Online Bounded Allocation Problem.

\noindent\rule{\textwidth}{0.5pt}
\end{spacing}
\vspace{5pt}
\begin{enumerate}[\text{Stage} 1:]
	\item As long as some interested buyers in $S_{j}$ spend less than $\eta$ fraction of their budget, the algorithm allocates item $j$ equally to buyers in $S_{j}$ who are on the lowest level.
\label{algo:bounded-allocation}
	\item Once all interested buyers spend at least $\eta$ fraction of their budgets, the algorithm assigns the remaining fraction of item $j$ to the predicted buyer $i^{*}$. The prediction assignment ends when one of the two following conditions occurs.
	\begin{enumerate}[(a)]
	    \item the algorithm assigned to the predicted buyer either $(1-\eta)$ fraction or the remaining fraction of $j$, whichever is smaller
	    \item buyer $i^{*}$ exhausted its budget
	\end{enumerate}
	\item If the algorithm does not assign item $j$ completely during Stage $1$ and $2$ and there exists at least one buyer in $S_{j}$ that does not exhaust its budget, then, similarly to the first step, the algorithm allocates the remaining fraction of item $j$ equally to buyers in $S_{j}$ who are on the lowest level.
\end{enumerate}
\begin{spacing}{0.1}
\noindent\rule{\textwidth}{0.5pt}
\end{spacing}

\subparagraph{Analysis.} We first prove the consistency and then the robustness of the algorithm.

\begin{lemma} \label{lem:bounded-allocation-consistency}
Algorithm \ref{algo:bounded-allocation} is $(1-\eta)$-consistent with the prediction.
\end{lemma}
\begin{proof}
Let $V$ be the set of buyers who exhaust their budget at some point during the algorithm's execution. Let $U$ be the set of items whose buyers did not exhaust their budgets, formally, $U =\{j : \pred(j) \notin V\}$. We can formulate the total gain of a feasible prediction as
\begin{equation} \label{eq:pred-gain}
\sum_{\genfrac{}{}{0pt}{2}{j}{\pred(j) \ne \emptyset}} b_{j}\ = \sum_{\genfrac{}{}{0pt}{2}{j}{\pred(j) \in V}} b_{j}\ + \sum_{\genfrac{}{}{0pt}{2}{j}{\pred(j) \notin V}} b_{j}
\ \le\ \sum_{i \in V}\ B_{i} + \sum_{j \in U}\ b_{j}
\end{equation}
where the inequality holds since any feasible allocation - including the prediction - can allocate items of value at most $B_{i}$ (the total budget) to buyer $i$.

Let $j \in U$ be an arbitrary item. By the construction of the algorithm and the fact that the predicted buyer for item $j$ did not exhaust its budget, the following can occur. The algorithm assigned item $j$ entirely during the limit assignment of Stage $1$ (to satisfy the condition of $\eta$ fraction spent by each interested buyer), or otherwise allocated the remaining fraction of item $j$ to its predicted buyer $i$ up to $(1-\eta)$ fraction. We note that the algorithm can always assign the remaining fraction, since buyer $i$ did not exhaust its budget by the end of the execution. In any case, we sell item $j$ in at least $(1 - \eta)$ fraction, therefore the gain of the algorithm on $j$ is at least $(1 - \eta) \ b_{j}$.

Before the prediction assignment of Stage $2$, the total value of items in $U$ sold to buyers in $V$ is at most $\sum_{i \in V} \eta B_{i}$ due to the limit assignment rule of Stage $1$. As a consequence, each predicted buyer $i^*$ where $i^* \notin V$ gets \emph{at least} $(1 - \eta) - \left(\left(\eta\ \sum_{i \in V}\ B_{i}\right) / b_j\right)$ fraction of item $j$. Therefore, the total gain of the algorithm is at least
\begin{equation}	\label{eq:algo-gain}
\sum_{i \in V}\ B_{i} + \sum_{j \in U}\ (1 - \eta) \ b_{j} - \eta\ \sum_{i \in V}\ B_{i}
\end{equation}
where the first term is the gain of the algorithm restricted to buyers in $V$, the second one is the gain of the algorithm restricted to items in $U$, and the last term is the upper bound of the total value of items in $U$ sold to buyers in $V$ before the prediction assignment of Stage $2$. By (\ref{eq:pred-gain}) and (\ref{eq:algo-gain}), the gain of the algorithm is at least $(1 - \eta)$ that of the prediction.
\end{proof}

Let us now establish the robustness of the algorithm.
Let $d = \max\{|S_j|\}$ be the \emph{bound} on the number of interested buyers.
We characterize the robustness as a function of $d$.
Let us define a piece-wise linear function $f_{d}: \left[0,1\right] \rightarrow \mathbb{R}_{\geq 0}$
such that $f_{d}(1) = 1$ and for $0 \leq u <1$, if $\frac{\ell - 1}{d} \leq u < \frac{\ell}{d}$ for some $1 \leq \ell \leq d$ then
\begin{align*}
	f_d \left( u \right) &= da_{\ell} \biggl( u - \frac{\ell - 1}{d} \biggr)
					+ da_{\ell - 1} \frac{1}{d} + da_{\ell - 2} \frac{1}{d} + \ldots + da_{1} \frac{1}{d} \\
				  &= da_{\ell} \biggl( u - \frac{\ell - 1}{d} \biggr)
					+ a_{\ell - 1}  + a_{\ell - 2} + \ldots + a_{1}
\end{align*}
where
$$
a_{1} = \frac{1}{d (1 + \frac{1}{d - 1})^{d-1} - (d - 1)}
\quad
\text{and}
\quad
a_{\ell} = a_{1} \biggl( 1 + \frac{1}{d-1} \biggr)^{\ell-1} \forall\ 2 \leq \ell \leq d.
$$
Informally, $f_{d}$ is linear with coefficient $da_{\ell}$
on every interval $\bigl[ \frac{\ell - 1}{d}, \frac{\ell}{d} \bigr)$ for $1 \leq \ell \leq d$.
Note that the maximum derivative of $f_d$ is $1/C(d)$, where
\[
C(d) = \frac{d \ (1 + \frac{1}{d-1})^{d-1} - (d-1)}{d \ (1 + \frac{1}{d-1})^{d-1}}=  1 - \frac{d-1}{d \ (1 + \frac{1}{d-1})^{d-1}}
	\overset{d \to \infty}{\longrightarrow} \frac{e-1}{e}
\]
and $C(d)$ is always larger than $(e-1)/e$.

\begin{lemma} \label{lem:bounded-allocation-robustness}
The robustness of Algorithm \ref{algo:bounded-allocation} is $1/\bigl( \frac{1}{C(d)} + (1-\eta)(1 - f_{d}(\eta)) \bigr)$.
When $d$ is large enough, $f_{d}(\eta) \approx 1 + \frac{e(e^{\eta - 1} - 1)}{e-1}$ and so the robustness is
approximatively $\frac{e-1}{e} \cdot \frac{1}{1 + (1-\eta)(1 - e^{\eta-1})}$.
\end{lemma}
\begin{proof}
We fix an arbitrary item $j$ and bound the ratio of increase of the primal and the dual objective values caused by the arrival of item $j$.
We set the dual variables as
\[y_i = f_d\left(\frac{\sum_{j} b_j x_{ij}}{B_i}\right) \quad \forall i,
\qquad
z_{j} = (1 - f_{d}(\min_{i \in S_{j}}\ \{\ell_{i}\}))\ b_{j}
\]
where $\ell_{i}$ is the level of buyer $i$ at the end of the algorithm's execution.

By the definition of the dual variables, $y_i \geq f_d\left(\min_{i' \in S_{j}}\ \{\ell_{i'}\}\right)$ for every $i \in S_{j}$.
Therefore, $b_{j} y_{i} + z_{j} \geq b_{j}$ holds for every $i \in S_{j}$, which indicates that the dual variables are feasible.

Let us assume that item $j$ is not entirely sold during the algorithm's execution.
This means that all interested buyers in $S_j$ exhausted their budget.
Hence, by the definition of the dual variables, $\forall\ i \in S_j : y_i = 1$ and $z_j = 0$.
The rate of change of the dual objective value related only to the $y$-variables is at most:
\[
B_{i} \frac{\partial f_d}{\partial x_{ij}}
	\leq B_{i} \cdot \frac{b_j}{B_{i}} \cdot f'_d\left(\frac{\sum_{j} b_j x_{ij}}{B_i}\right)
	\leq  \frac{b_j}{C(d)}
\]
where $1/C(d)$ is the maximum derivative of $f_d$. Meanwhile, the increasing rate of the primal objective value is $b_{j}$. We obtain that the primal change is at least $1/C(d)$ that of the dual.

In the remaining part of the proof, we assume that the algorithm sold all items completely. Note that $z_{j}$ can be written as
$z_{j} = \int_{0}^{1} (1 - f_{d}(\min_{i \in S_{j}} \{\ell_{i}\})) \ b_{j}\ dy$. In other words, one can imagine that during the allocation of item $j$,
$z_{j}$ is increasing at a rate of $(1 - f_{d}(\min_{i \in S_{j}}\ \{\ell_{i}\}))\ b_{j}$.

There are three stages in the algorithm. In Stage 1 and 3, the algorithm always allocates the fractions of item $j$ equally
to buyers in $S_{j}$ on the lowest level. \citet{BuchbinderJain07:Online-primal-dual} showed that during these allocations, the increasing rate of the dual is at most $1/C(d)$ times the primal. We present our proof in Lemma~\ref{lem:bound-stages} for completeness, which is similar to the proof of \cite{BuchbinderJain07:Online-primal-dual}.

We are now interested in the allocation during Stage 2 of the algorithm. We denote the predicted buyer $i^{*} = \pred(j)$.
In this stage, the algorithm allocates a part of item $j$ only to the predicted buyer $i^{*}$.
The increasing rate of the dual objective value related only to $y_{i^*}$ is
\[
B_{i^*} \frac{\partial f_d}{\partial x_{i^*j}}
	\leq B_{i^*} \cdot \frac{b_j}{B_{i^*}} \cdot f'_d\left(\frac{\sum_{j} b_j x_{i^{*}j}}{B_i^{*}}\right)
	\leq  \frac{b_j}{C(d)}
\]
since $f'_{d}(u) \leq 1/C(d)$ for all $0 \leq u \leq 1$.
We note that every buyer $i \in S_{j}$ has spent at least $\eta$ fraction of its budget before this stage, so $z_{j} \leq (1 - f_{d}(\eta))\ b_{j}$.
Therefore, the total increasing rate of the dual in this step is
at most
\[
B_{i^*} \frac{\partial f_d}{\partial x_{i^*j}}  + \frac{\partial z_{j}}{\partial x_{i^*j}}
	\le \frac{b_{j}}{C(d)} + (1 - f_d(\eta))\ b_{j}.
\]
Combining all the cases, the increasing rate of the dual is at most
\begin{align*}
\begin{cases}
\frac{b_{j}}{C(d)} + (1 - f_d(\eta))\ b_{j} & \textnormal{during Stage 2}, \\
\frac{b_{j}}{C(d)} & \textnormal{during Stage 1 and 3}.
\end{cases}
\end{align*}
We highlight that the algorithm allocates at most $(1 - \eta)$ fraction of item $j$ during Stage 2.
Therefore, the total increase of the dual due to the arrival of $j$ is at most:
$$
\int_{0}^{(1-\eta)} \biggl( \frac{b_{j}}{C(d)} + (1 - f_d(\eta))\ b_{j}  \biggr)\ dy + \int_{0}^{\eta} \biggl( \frac{b_{j}}{C(d)} \biggr)\ dy
= \biggl( \frac{1}{C(d)} + (1-\eta)(1 - f_{d}(\eta)) \biggr) b_{j}.
$$
The increase of the primal is $b_{j}$, so we can deduce that the robustness of the algorithm is
$1/\bigl( \frac{1}{C(d)} + (1-\eta)(1 - f_{d}(\eta)) \bigr)$. To finish the proof, we compute and estimate the value of $f_d(\eta)$ for large values of $d$. We have
\begin{align*}
f_{d}(\eta) = f_d\biggl(\frac{ \lfloor \eta \cdot d \rfloor}{d}\biggr) &= a_1\ \biggl(d \left(1 + \frac{1}{d-1}\right)^{\lfloor \eta \cdot d \rfloor -1} - (d-1)\biggr) \\
&= \frac{d(1 + \frac{1}{d-1})^{\lfloor \eta \cdot d \rfloor - 1} - (d-1)}{d(1 + \frac{1}{d-1})^{d-1} - (d-1)}\\
&= 1 + \frac{d(1 + \frac{1}{d-1})^{\lfloor \eta \cdot d \rfloor - 1} - d(1 + \frac{1}{d-1})^{d-1}}{d(1 + \frac{1}{d-1})^{d-1} - (d-1)} \\
&= 1 + \left(\frac{d(1+\frac{1}{d-1})^{\lfloor \eta \cdot d \rfloor -1}}{d(1+\frac{1}{d-1})^{d-1}} - 1\right) \cdot \frac{1}{1 - \frac{(d-1)}{d(1 + \frac{1}{d-1})^{d-1}}} \\
&= 1 + \frac{1}{C(d)} \left(\left(1+\frac{1}{d-1}\right)^{\lfloor \eta \cdot d \rfloor - d} - 1\right) \\
&\overset{d \to \infty}{\longrightarrow}
1 + \frac{e(e^{\eta-1}-1)}{e-1}
\end{align*}
\end{proof}

\begin{lemma} \label{lem:bound-stages}
Assuming that item $j$ is completely sold by Algorithm~\ref{algo:bounded-allocation}, during the allocations of Stage 1 and 3, the increasing rate of the dual objective value is at most $1/C(d)$ times the primal.
\end{lemma}

\begin{proof}
The proof follows the proof of \citet[Theorem 13.1]{BuchbinderNaor09:The-Design-of-Competitive}.
We present the details in \cref{apix:bounded-allocation}.
\end{proof}

\setcounter{theorem}{0}
\begin{theorem} \label{theorem:bounded-allocation}
Algorithm \ref{algo:bounded-allocation} is $(1-\eta)$-consistent and $1/\bigl( \frac{1}{C(d)} + (1-\eta)(1 - f_{d}(\eta)) \bigr)$-robust.
When $d$ is large enough, $f_{d}(\eta) \approx 1 + \frac{e(e^{\eta - 1} - 1)}{e-1}$ and so the robustness is
approximately $\frac{e-1}{e} \cdot \frac{1}{1 + (1-\eta)(1 - e^{\eta-1})}$.
\end{theorem}
\begin{proof}
Follows from Lemma \ref{lem:bounded-allocation-consistency} and Lemma \ref{lem:bounded-allocation-robustness}.
\end{proof}

\section{An Algorithm with Predictions for Ad-Auctions} \label{sec:ad-auction}

Recall that in this problem there are $n$ buyers and each buyer $1 \leq i \leq n$ has a budget $B_{i}$. Upon the arrival of item $j$,
the algorithm discovers the bid\footnote{we call the prices bids because of the motivations in the auctions setting}
$b_{ij} \geq 0$ of each buyer $i$, which is the price that buyer $i$ is willing to pay to purchase item $j$. Additionally, the algorithm gets a predicted buyer $\pred(j)$ to whom item $j$ should be sold ($\pred(j) = 0$ if the item should not be sold according to the prediction).
We are interested in fractional solutions as in \citet{BamasMaggiori20:The-Primal-Dual-method}, so we consider the items to be splittable. Before the arrival of the next item, the algorithm needs to make an \emph{irrevocable} decision
and sell the current item in some fractions to some buyers.
The objective is to gain maximum revenue from selling items to buyers. In this setting, we assume that $b_{ij} \ll B_{i}~\forall i,j$.


\subparagraph{Formulation.} The formulation of the Online Ad-Auctions problem follows that of the Online Bounded Allocation problem. Both the primal and the dual linear programs on \cref{fig:formulation2} now use the buyer-dependent prices, the bids.

\begin{figure}[h!]
\centering
\begin{minipage}[t]{0.42\textwidth}
Primal:
\begin{align*}
\max  \sum_{i=1}^{n} \sum_{j=1}^{m} &\ b_{ij}\ x_{ij} \\
\sum_{j=1}^{m}\ b_{ij} \ x_{ij} &\leq B_i && \forall\ i & (y_i)\\
\sum_{i=1}^{n}\ x_{ij} &\leq 1 && \forall\ j &  (z_j)\\
x_{ij} &\geq 0 && \forall\ i,\ j & \\
\end{align*}
\end{minipage}
\quad
\begin{minipage}[t]{0.5\textwidth}
Dual:
\begin{align*}
\min\ \sum_{i=1}^{n}\ B_i\ y_i &+ \sum_{j=1}^{m}\ z_j  \\
 b_{ij} \ y_i + z_j &\ge b_{ij} && \forall\ i,\ j & (x_{ij})\\
y_i &\ge 0 && \forall\ i & \\
z_j &\ge 0 && \forall\ j & \\
\end{align*}
\end{minipage}
\caption{Formulation of the Online Ad-Auctions problem}
\label{fig:formulation2}
\end{figure}

%

\subparagraph{Algorithm.}
We introduce a fictitious buyer with identity $0$, such that $b_{0j} = 0$ for all item $j$. When the algorithm does not sell an item, it assigns it to the fictitious buyer $0$. The purpose of buyer $0$ is to simplify the description of the algorithm. We use a constant $C$ in our algorithm that we define as
\[C = (1 + R_{\max})^{\frac{\eta}{R_{\max}}} \ \ \textnormal{ where } R_{\max} = \max_{i,j} \left\{\frac{b_{ij}}{B_{i}}\right\}\]
We present the pseudo-code of our algorithm in \Cref{algo:ad-auctions}. For each arriving item $j$, the algorithm considers two buyers: the predicted buyer $i^* = \pred(j)$ and buyer $i$ who maximizes the product $b_{ij}~(1-y_j)$, where $y_j$ is an indicator to know how exhausted the buyer's budget is.
We make use of the confidence parameter in the predictions.
Our algorithm reserves $(1-\eta)$ fraction of each buyer's budget for the prediction assignment. Whenever the bid of buyer $i^*$ is greater than the bid of buyer $i$, the algorithm assigns $(1-\eta)$ fraction of item $j$ to $i^*$ and $\eta$ fraction to $i$.

Intuitively, our proposed algorithm attempts to reserve some fraction of each buyer's budget for future purchases. However, if the predicted buyer's bid is high enough, the algorithm allows assignments to this buyer, even if the budget is close to saturation.

\setcounter{algorithm}{1}
\begin{algorithm}[ht]
\DontPrintSemicolon
All primal and dual variables are initially set to 0.\\
We maintain two sets $N(i)$ and $M(i)$ for each buyer $i$ for the purpose of analysis only.

\For{each new item $j$} {
    let $i^*$ be the predicted buyer of item $j$, formally, $\pred(j) = i^*$\\
    \textbf{if} the prediction is not feasible \textbf{then} $i^* = 0$ \tcp*{\small the fictitious buyer}
    $i \gets \arg\max_{i'}\ \{b_{i'j}\ (1-y_{i'})\}$ \tcp*{\small weight bids with remaining budget}
    \textbf{if} $b_{ij}\ (1 - y_i) \leq 0$ \textbf{then} $i = 0$

    $z_j \gets \max \bigl \{0,\  b_{ij} (1 - y_i) \bigr \}$

    \If{$b_{ij} < b_{i^{*}j}$}{
        $x_{ij} \gets \eta$ and $x_{i^{*}j} \gets (1-\eta)$

        $N(i^{*}) \gets N(i^{*}) \cup \{j\}$
    } \Else {
        $x_{ij} \gets 1$ \tcp*{\small includes the case when $\pred(j)$ is infeasible}
    }

    $M(i) = M(i) \cup \{j\}$

    $y_i = y_i \left(1 + \frac{b_{ij}}{B_{i}} \right) + \frac{b_{ij}}{B_{i}} \cdot \frac{1}{C - 1}$
}
\caption{Learning Augmented Algorithm for the Online Ad-Auctions Problem.}
\label{algo:ad-auctions}
\end{algorithm}


\setcounter{theorem}{3}
\begin{lemma}\label{lem:alpha}
During the execution of Algorithm \ref{algo:ad-auctions} the following always holds for every $i$
\[y_i \geq \frac{1}{C - 1} \biggl( C^{\frac{\sum_{j \in M(i)} b_{ij}}{\eta B_{i}}} - 1 \biggr)\]
\end{lemma}
\begin{proof}
We adapt the proof of \cite{BuchbinderNaor09:The-Design-of-Competitive} with a slight modification: we prove the dual inequality by induction on the number of processed items. Initially, when no items arrived yet, the inequality is trivially true. Let us assume that the inequality holds right before the arrival of an item $j$. The inequality remains unchanged for all buyers, except for buyer $i$, who maximizes $b_{ij} \ (1 - y_i)$. Let $y_i$ denote the value before the update triggered by the arrival of item $j$ and $y'_i$ its value after the update.
We have
\begin{align*}
y'_{i}
= y_{i}  \left( 1 + \frac{b_{ij}}{B_{i}} \right)
										&+  \frac{b_{ij}}{B_{i}} \cdot \frac{1}{C - 1}
\geq \frac{1}{C - 1} \biggl( C^{\frac{\sum_{j' \in M(i) \setminus \{j\} } b_{ij'}}{\eta B_{i}}} - 1 \biggr)
			\cdot \left( 1 + \frac{b_{ij}}{B_{i}} \right)
										+   \frac{b_{ij}}{B_{i}} \cdot \frac{1}{C - 1} \\
&=  \frac{1}{C - 1} \biggl( C^{\frac{\sum_{j' \in M(i) \setminus \{j\} } b_{ij'}}{\eta B_{i}}}
						\cdot \left( 1 + \frac{b_{ij}}{B_{i}} \right)  - 1 \biggr) 	\\
&\geq  \frac{1}{C - 1} \biggl( C^{\frac{\sum_{j' \in M(i) \setminus \{j\}} b_{ij'}}{\eta B_{i}}}
						\cdot  C^{\frac{b_{ij} }{\eta B_{i}}}   - 1 \biggr)
= \frac{1}{C - 1} \biggl( C^{\frac{\sum_{j \in M(i)} b_{ij}}{\eta B_{i}}} - 1 \biggr)
\end{align*}
The first inequality holds due to the induction hypothesis. The second inequality holds by the following sequence of transformations. For any $y$ and $z$ where $0 < y \leq z \leq 1$ we have
\begin{align*}
\frac{\ln(1+y)}{y} \geq \frac{\ln(1+z)}{z}
\quad \Leftrightarrow \quad \ln(1+y) \geq \ln(1+z)\ \cdot \frac{y}{z}
\quad \Leftrightarrow \quad 1+y \geq (1+z)^{y/z}
\end{align*}
We apply the above transformation with $y = b_{ij}/{B_i}$ and $z = R_{\max}$ at the second inequality. By the definition of $C$ we obtain
\[
\biggl( 1+R_{\max} \biggr)^{\frac{1}{R_{\max}} \cdot \frac{b_{ij} }{B_{i}}}
= C^{\frac{b_{ij}}{\eta B_{i}}}
\]
Since $R_{\max} = \max_{i,j} \left\{\frac{b_{ij}}{B_{i}}\right\}$ by definition, the induction step is complete and the lemma holds.
\end{proof}

\begin{lemma}	\label{lem:ad-auctions-primal-feasibility}
The primal solution is feasible up to a factor of $(1 + R_{\max})$.
\end{lemma}
\begin{proof}
The first primal constraint requires $\sum_{j=1}^{m} b_{ij}\ x_{ij} \leq  B_{i}$ to hold for every $i$. By Lemma~\ref{lem:alpha}, we know that for every $i$ the following holds.
\[
y_{i} \geq \frac{1}{C - 1} \biggl( C^{\frac{\sum_{j \in M(i)} b_{ij}}{\eta B_{i}}} - 1 \biggr)
\]
Therefore, whenever $\sum_{j \in M(i)} b_{ij} \geq \eta B_{i}$, we have $y_{i} \geq 1$ and the algorithm stops allocating items to buyer $i$. The set $M(i)$ and the value of $y_i$ are updated after the assignments. Therefore, buyer $i$ can receive at most one additional item fraction once its budget is already saturated. We obtain $\sum_{j \in M(i)} b_{ij}  < \eta B_{i} + \max_{j} \{b_{ij}\}$ and the following formula.
\begin{align*}
\sum_{j=1}^{m} b_{ij}\ x_{ij} =
\sum_{j \in M(i)} b_{ij}\ x_{ij} + \sum_{j \in N(i)} b_{ij}\ x_{ij}
<  B_{i} + \max_{j}\ \{b_{ij}\}
\end{align*}
The inequality holds due to the feasibility of $N(i)$, the set of items assigned by the prediction to buyer $i$. We can bound the prediction assignments as follows.
$$
\sum_{j \in N(i)} b_{ij}\ x_{ij} \leq  (1 - \eta) \sum_{j\ |\ \pred(j) = i} b_{ij}\ x_{ij} \leq (1 - \eta)\ B_{i}
$$
Therefore,
$
\sum_{j=1}^{m} b_{ij}\ x_{ij} \leq B_{i}\ (1 + R_{\max})
$, which means that the first primal constraint is feasible up to a factor of $(1+R_{\max}$). The second primal constraint requires $\sum_{i} x_{ij} \leq 1$ to hold. During the allocations of \Cref{algo:ad-auctions} the values of $x_{ij}$ and $x_{i^{*}j}$ do not exceed $1$ by the design of the algorithm. The lemma follows.
\end{proof}

\setcounter{theorem}{1}
\begin{theorem} \label{theorem:ad-auction}
Algorithm \ref{algo:ad-auctions} is $(1 - \eta)$-consistent and $\frac{1 - 1/C}{1 + R_{\max}}$-robust.
The robustness tends to $1 - e^{-\eta}$ when $R_{\max}$ tends to 0.
\end{theorem}
\begin{proof}
First, we establish robustness. Upon the arrival of item $j$, the increase in the primal is
\[
\begin{cases}
	(1-\eta) \ b_{i^{*}j} + \eta \ b_{ij} & \text{ if } b_{ij} < b_{i^{*}j}, \\
	b_{ij} & \text{ if } b_{ij} \geq b_{i^{*}j}
\end{cases}
\]
which is always larger than or equal to $b_{ij}$. Meanwhile, the increase in the dual is
\begin{align*}
B_{i} \Delta y_{i}  + z_{j}
&= b_{ij} y_{i} + \frac{b_{ij}}{C-1}
 + b_{ij} (1 - y_{i})
 = \biggl( 1 + \frac{1}{C-1} \biggr) b_{ij}
 = \frac{C}{C - 1}\ b_{ij}
 \end{align*}
Hence, by Lemma~\ref{lem:ad-auctions-primal-feasibility}, the robustness is
$\frac{C-1}{C} \cdot \frac{1}{1+R_{\max}}$.

Finally, we straightforwardly establish consistency. Every time the prediction solution gets a profit of $b_{i^{*}j}$, the algorithm achieves a profit of
at least $(1 - \eta) \ b_{i^{*}j}$. The theorem follows.
\end{proof}

\section{Experiments}

In the following experiments, we evaluate the algorithms' performance with the competitive ratio metric. We calculate the competitive ratio by dividing the objective value of the algorithm, ALGO(I), with the objective value of the optimal fractional offline solution, OPT(I). The competitive ratio is visible on the $y$-axis of the figures. The $x$-axis corresponds to the ratio with which the algorithms consider the prediction. We indicate with $\eta = 0$ no doubt and $\eta = 1$ complete doubt in the prediction. Therefore, the left-hand side of the figures corresponds to a prediction dominated algorithm, while the right-hand side is closer to a classical online algorithm. The different colors on the figures correspond to different prediction error rates. The lines represent the average over several executions, while the colored areas show the $95\%$ confidence intervals.

\subsection{Online Bounded Allocation }

We present here four experiments. \cref{tab:instances} below summarizes the configurations of the instances. \cref{fig:experiment-bounded-allocation} displays the competitive ratio of the algorithm on different instances.

\textbf{Predictions.} We create predictions for the experimental instances by introducing perturbations to the optimal offline integral solution. The perturbation is skipped when $|S_j| = 1$. When $|S_j \setminus \{i^*\}| > 1$ we choose uniformly at random a remaining buyer to replace the optimal buyer.

\begin{center}
    \begin{tabular}{|c|c|r|r|c|c|c|}
        \hline
        Name & Type & Buyers & Items & $d = \max\{|S_j|\}$ & Budget range & Item price range \\
        \hline
        Instance $1$ & manual & $5$ & $5$ & $5$ & $100 - 100$ & $100 - 100$\\
        Instance $2$ & random & $100$ & $1.000$ & $5$ & $10 - 100$ & $0.1 - 8$ \\
        Instance $3$ & random & $100$ & $10.000$ & $3$ & $10 - 1.000$ & $1 - 10$ \\
        Instance $4$ & random & $80$ & $80$ & $40$ & $10 - 100$ & $10 - 100$\\
        \hline
    \end{tabular}
    \captionof{table}{Properties of the experiment instances}
    \label{tab:instances}
\end{center}

\subparagraph{Manual Instance.} Instance $1$ corresponds to one of the pathological inputs for the water-filling strategy of \cite{BuchbinderNaor09:The-Design-of-Competitive}. We created this instance manually to observe the behavior of Algorithm~\ref{algo:bounded-allocation} when this strategy performs poorly. In Instance $1$'s scenario, each buyer $i$ is interested in each item $j$ when $i \ge j$. The optimal solution is to sell each item $j$ to buyer $i$, where $i = j$, while the water-filling strategy attempts to allocate each item equally.

\subparagraph{Randomized Instances.} Instances $2$-$4$ are randomly generated based on their corresponding configuration visible on \cref{tab:instances}. We executed each instance $20$ times. The lines on \cref{fig:experiment-bounded-allocation} correspond to the average of these executions, and the colored areas correspond to the $95\%$ confidence intervals. Instance~$2$ and Instance~$3$ mimic real-life instances, where the general expectation is to have a small bound $d$ on the interested buyers. The average item price over the average budget value across the executions was $6.36\%$ for Instance~$2$ and $0.98\%$ for Instance~$3$. Finally, Instance~$4$ shows an example where the integral solution (and therefore the prediction) is not optimal. While Instance $2$ and $3$ have no integrality gap, Instance $4$ has an average of $17.99\%$ observed integrality gap over the $20$ executions. Instance~$4$ has a large bound on the number of interested buyers, and the items' prices vary greatly.

\begin{center}
  \minipage{0.25\textwidth}
    \centering
    \includegraphics[width=\linewidth]{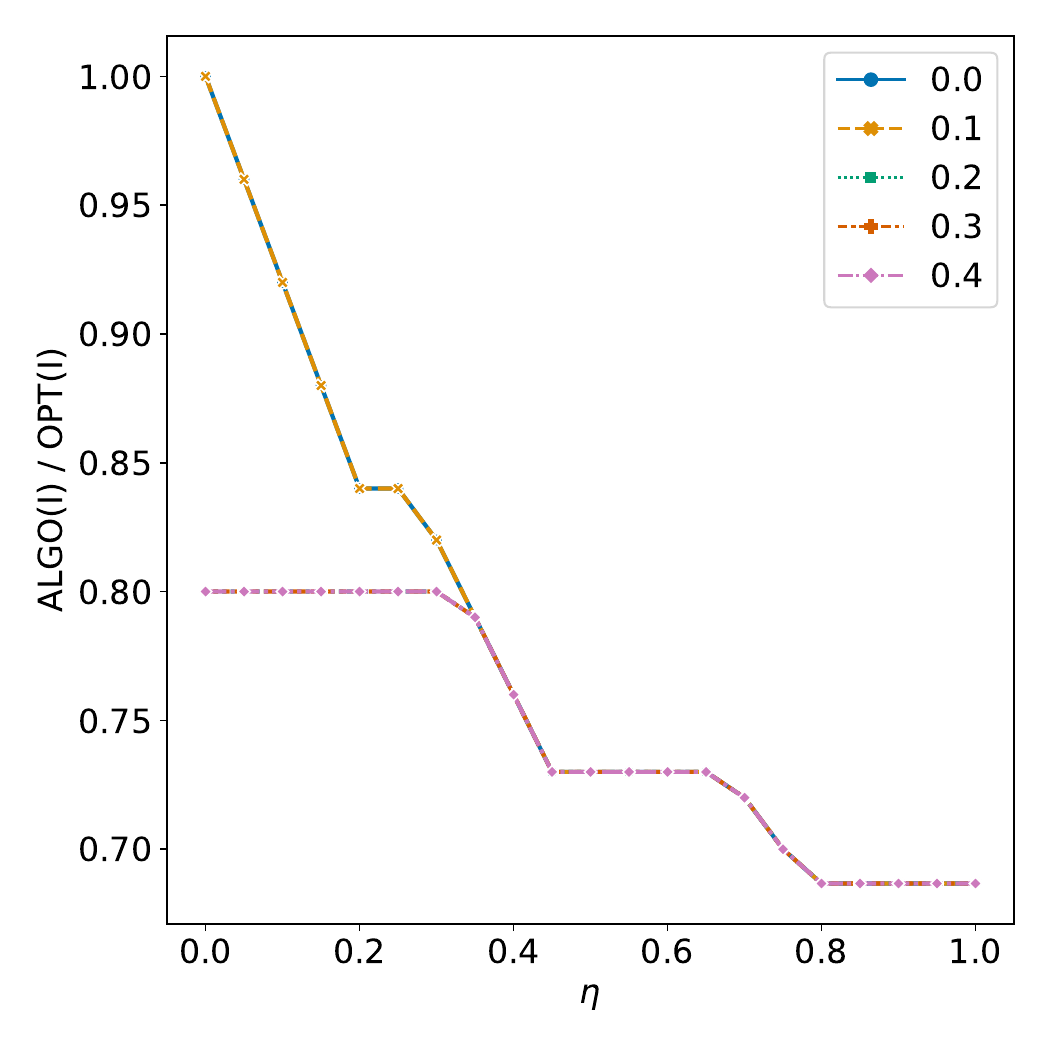}
    {Instance $1$}
    \label{fig:instance1}
  \endminipage\hfill
  \minipage{0.25\textwidth}
    \centering
    \includegraphics[width=\linewidth]{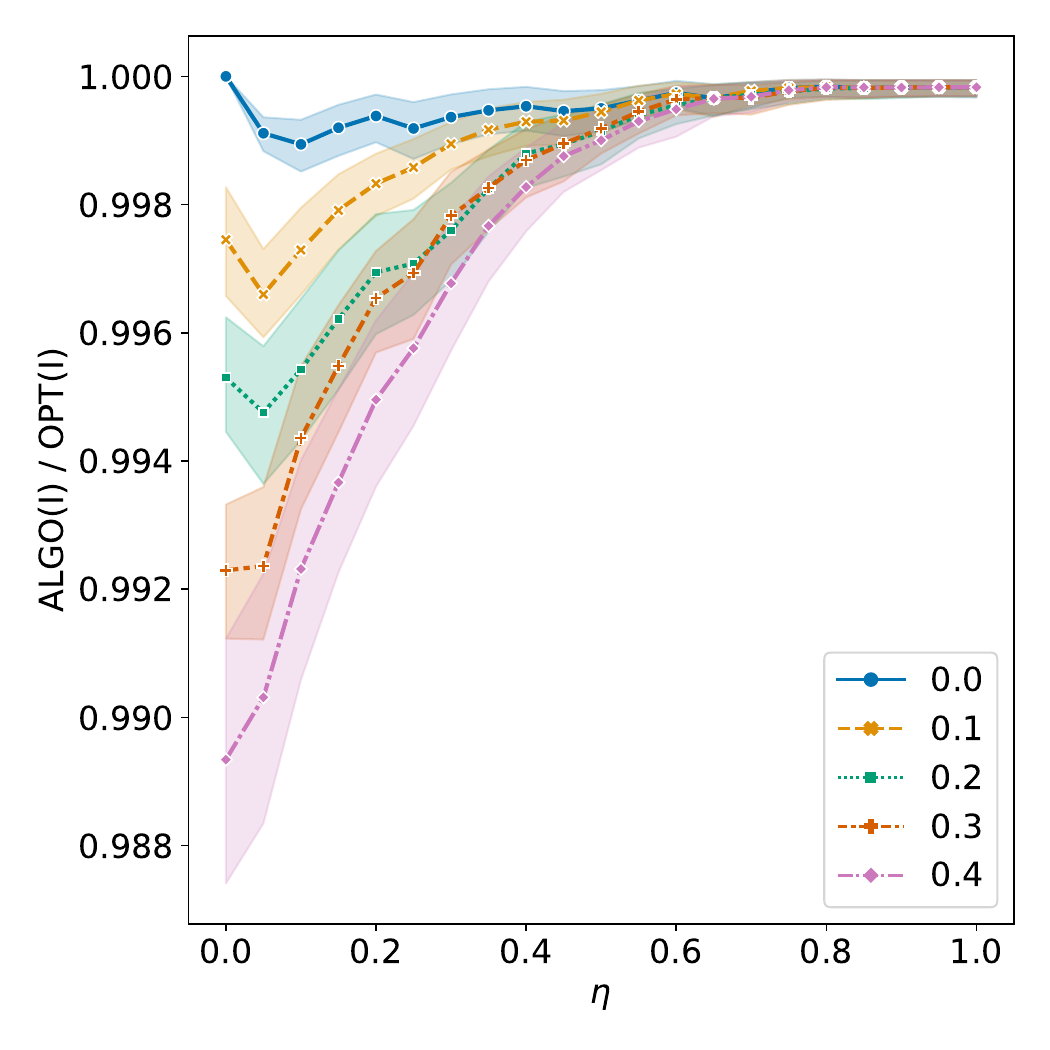}
    {Instance $2$}
    \label{fig:instance2}
  \endminipage\hfill
  \minipage{0.25\textwidth}
    \centering
    \includegraphics[width=\linewidth]{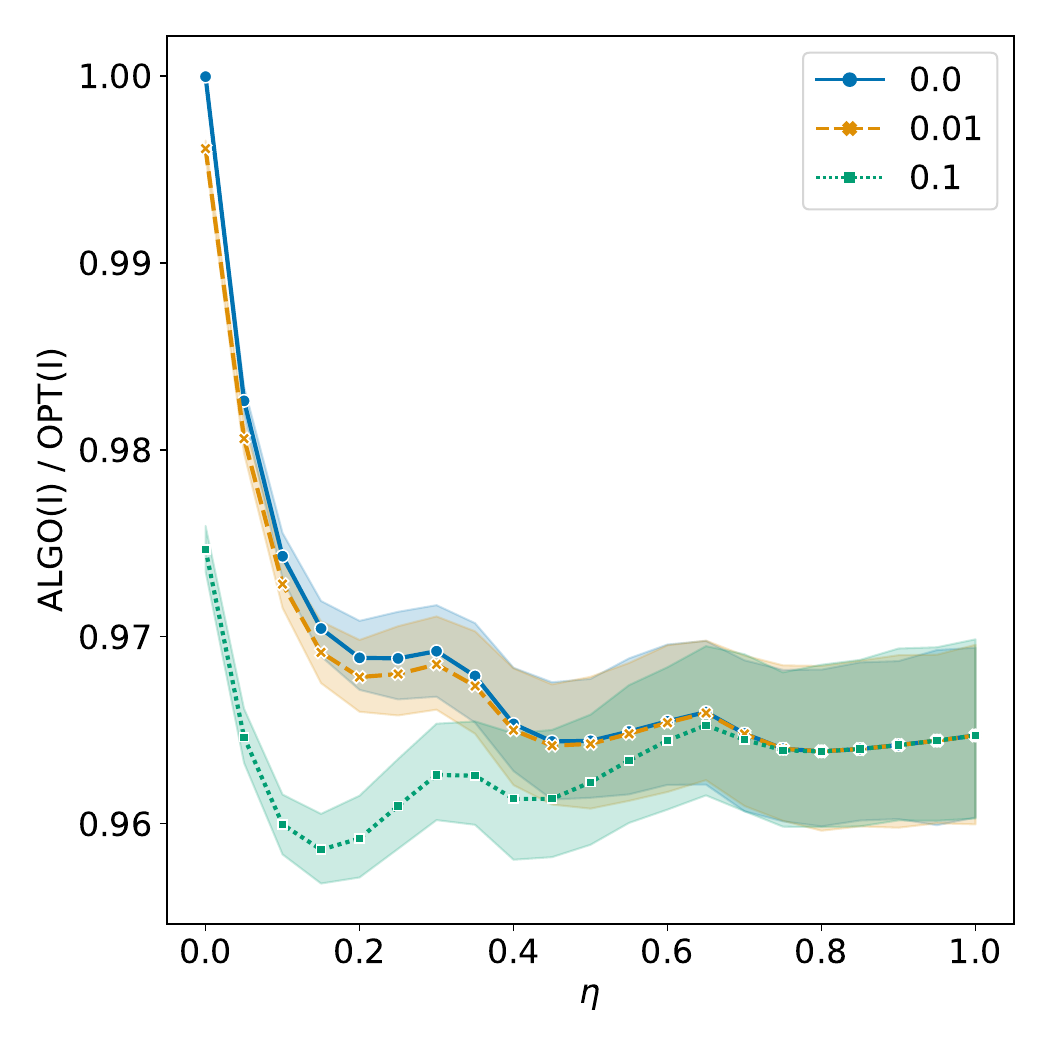}
    {Instance $3$}
    \label{fig:instance3}
  \endminipage\hfill
  \minipage{0.25\textwidth}
    \centering
    \includegraphics[width=\linewidth]{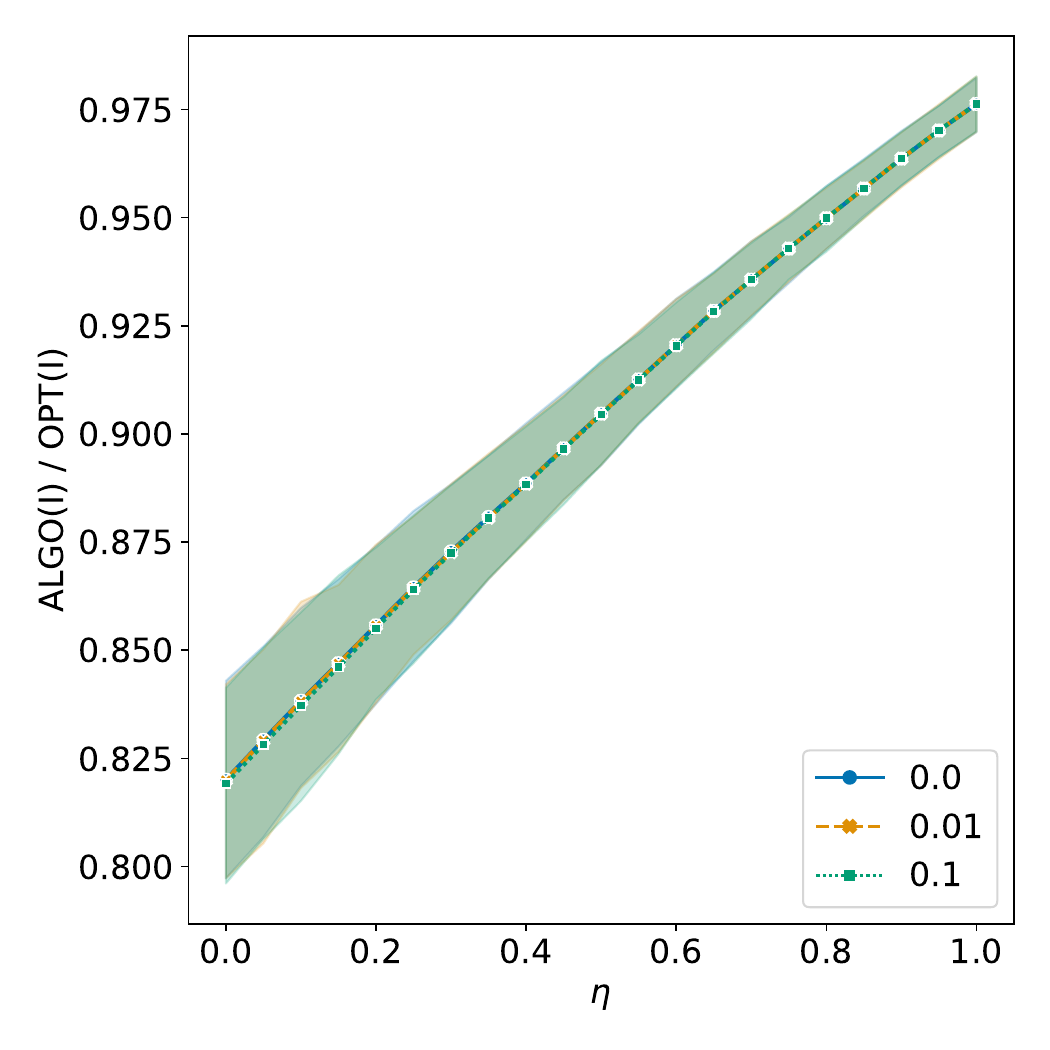}
    {Instance $4$}
    \label{fig:instance4}
  \endminipage\hfill

\hfill

\captionof{figure}{Observed competitive ratios}
\label{fig:experiment-bounded-allocation}
\end{center}

\subparagraph{Evaluation.} The experiments confirm the water-filling algorithm's benefit from the prediction information. We showed that Algorithm~\ref{algo:bounded-allocation} has improved performance on the pathological input of the water-filling strategy even with high prediction error rates. Besides, the algorithm demonstrated firm robustness against elevated prediction error rates in the second instance, where the algorithm performed close to optimality without predictions. Furthermore, we observed improved performance on the third instance, which represents a close to real-life use case. In the fourth instance, we can remark the drawback of the predictions. When the integral solution of the linear program is not optimal - or even far from optimal - the prediction can significantly misguide our algorithm, decreasing the performance towards the standard performance bounds.

\subsection{Online Ad-Auctions}

We present one experiment for the Online Ad-Auctions problem here. \cref{fig:ad-auction-experiments} displays the competitive ratio of the algorithm.

\subparagraph{Predictions.} The predictions rely on the optimal offline integral solution, which is a partial mapping from items to buyers. We perturbed the solution as suggested by \cite{BamasMaggiori20:The-Primal-Dual-method} and used the error rate parameter as a probability to randomly choose a buyer
among the buyers who placed a non-zero bid on the item. However, the perturbation is only possible if the solution remains feasible.

\begin{multicols}{2}
\subparagraph{Instance.} Our test instance is a randomized instance with $100$ buyers and $10,000$ items, adapting the model described in \cite{Lavastida20:Predictions-Matching}. For every item $j$, there are exactly $6$ random buyers proposing a bid. The values of the bids follow a $lognormal$ distribution with mean and deviation set to $0.5$. By choosing the budget of the bidders, we can tune the hardness of the instance under the constraint that $R_{\max}$ remains reasonably small. We set the budget to $0.1$ fraction of the total bids, leading to a value of $R_{\max} \approx 0.1$. The integrality gap of the instance is close to $0$.

\begin{center}
\includegraphics[width=0.73\linewidth]{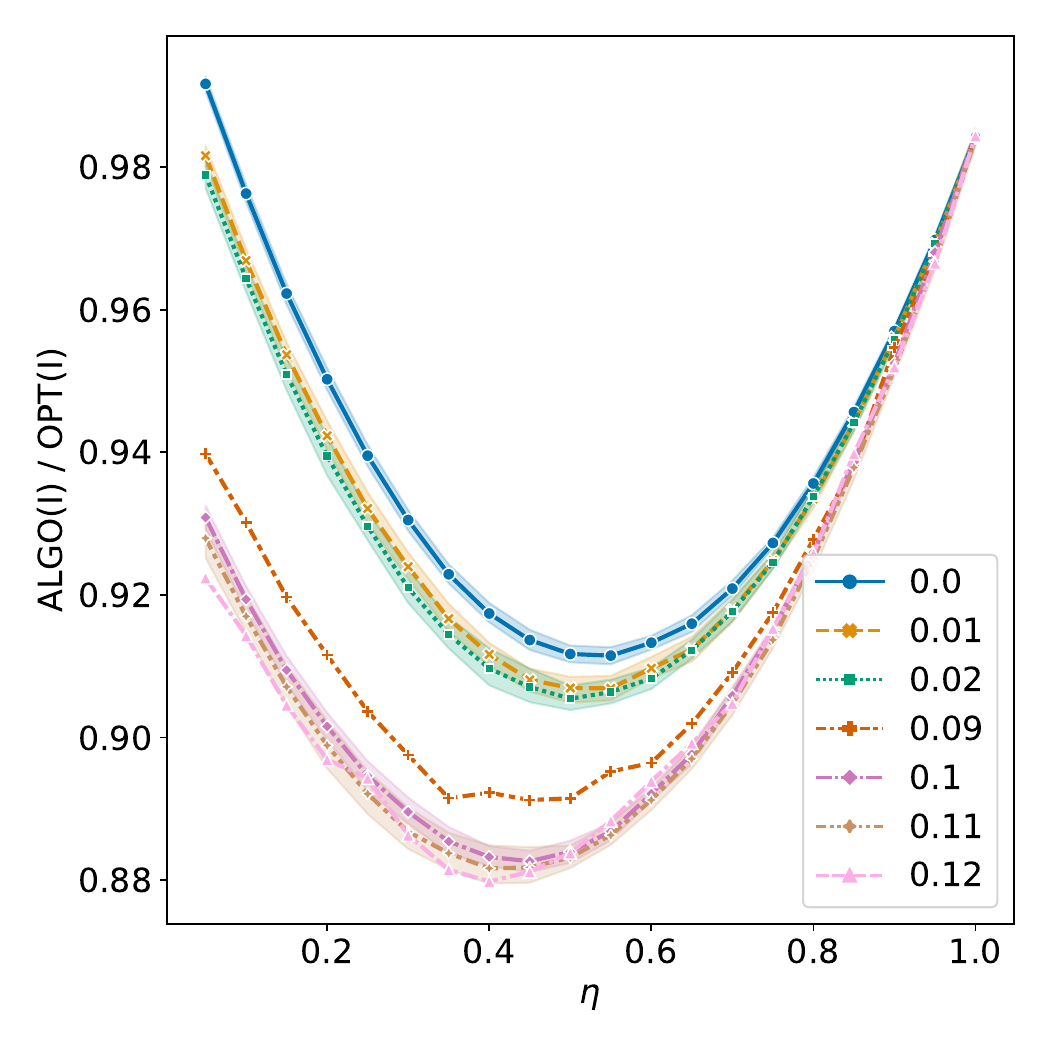}
\captionof{figure}{Observed competitive ratio}
\label{fig:ad-auction-experiments}
\end{center}

\hfill
\end{multicols}

\subparagraph{Evaluation.} We can observe the algorithm's benefit of the prediction. We note that in the algorithm $\eta$ can not be $0$ since, in this case, the update of $y_i$ would require division by $0$. The closer $\eta$ gets to $0$, the closer the performance of the algorithm gets to $1$. As expected, the performance degrades with increased prediction perturbation. The observed robustness is not monotone in $\eta$, unlike the bound shown in this paper. We think that this performance degradation for $\eta$ around $0.5$ is due to the rather simplistic mixture between the primal-dual and the predicted solution.

\section{Conclusion}
The presented algorithms in this paper incorporate prediction information and achieve competitive consistency and robustness. However, it remains an open question to determine lower bounds and verify how tight our obtained bounds are.

An important research direction for the future is to study matching problems with dynamic confidence parameters, which change over time depending on the quality of the previous predictions. In general, it might be possible to design more complete systems that evolve through the interactions between the algorithms and the learning oracles. The oracles provide useful information (predictions) for the algorithms to improve their performance (as studied in this paper), while the algorithms could give feedback to the oracles to enhance their predictions.

\acks{This research has been partly supported by the research
program on Edge Intelligence at the Multidisciplinary
Institute on Artificial Intelligence MIAI in Grenoble (ANR-19-P3IA-0003) and by the french research agency Energumen (ANR-18-CE25-0008).}

\bibliography{references}

\clearpage

\appendix

\section{for the Online Bounded Allocation problem} \label{apix:bounded-allocation}

\begin{lemma}
Assuming that item $j$ is completely sold by Algorithm~\ref{algo:bounded-allocation}, during the allocations of Stage 1 and 3, the increasing rate of the dual objective value is at most $1/C(d)$ times the primal.
\end{lemma}

\begin{proof}
The proof follows the proof of \citet[Theorem 13.1]{BuchbinderNaor09:The-Design-of-Competitive}.
We highlight that this proof considers items that were sold completely. There are two cases.

\textit{Case 1}: The highest level item $j$ was sold at is $\ell$, and at the end of the algorithm, all buyers in $S_j$ spent at least $\frac{\ell+1}{d}$ fraction of their budget.

By the definition of the dual variables $z_j \leq (1 - f_d\bigl( \frac{\ell+1}{d} \bigr)) \ b_j$. Since item $j$ is sold entirely ($\sum_{i} x_{ij} = 1$), the increasing rate of $z_{j}$ at any time during the allocation of item $j$ is at most $(1 - f_d\bigl( \frac{\ell+1}{d} \bigr))\ b_j$.
Besides, as the highest level on which item $j$ was sold at is $\ell$, the rate of change of the dual value due to the change in $y_i$ is:
\[
B_i\ \frac{\partial f_d}{\partial x_{ij}} \le B_i \ \frac{b_j}{B_i} \ d \ a_{\ell+1} \le b_j \ d \ a_{\ell+1}
\]
Therefore, the total change of the dual is at most
\begin{align}	\label{eq:allocation-case1}
 &b_j\ d\ a_{\ell+1} + b_j \biggl(1 - f_d\biggl( \frac{\ell+1}{d} \biggr) \biggr) \notag \\
 &= b_{j}\ d\ a_{1} \left( 1 + \frac{1}{d-1}  \right)^{\ell} + b_{j} \left( 1 - a_{1} \biggl( d \biggl( 1 + \frac{1}{d-1} \biggr)^{\ell} - (d-1)  \biggr) \right) \notag \\
 &= b_{j} (1 + a_{1}(d-1))
 = \frac{b_j}{C(d)}
 \end{align}

\textit{Case 2:} The highest level on which item $j$ was sold at is $\ell$ and at the end of the execution at least one buyer in $S_{j}$ spent less than $\frac{\ell+1}{d}$ fraction of its budget.

In this case we have to set $z_j = (1 - f_d(\ell/d)) \ b_j$ to satisfy the dual constraint.
During any (short) period of time $\Delta t$, the increase of $\sum_{i} B_{i}\ y_{i}$ over $i$ in the highest level is at most:
\[
b_j \ a_{\ell+1} \ d \ \frac{d-1}{d} \ \Delta t = b_j\ (d-1)\ a_{\ell+1}\ \Delta t
\]
where the fraction $(d-1)/d$ comes from the fact that there are at most $(d-1)$ buyers in the highest level. We recall that $d$ is the bound on the size of all $S_j$ and at least one buyer did not reach the highest level as of the statement of this case. The increase of $\sum_{i} B_{i}\ y_{i}$ over $i$ in the lower levels is at most $b_j\ d\ a_{\ell}\ (1 - \Delta t)$ by the definitions of $a_{\ell+1} = a_\ell(1+1/(d-1))$ and $a_\ell = ((d-1)/d)a_{\ell+1}$. Therefore, by adding the terms together we get that the increasing rate of $\sum_{i} B_{i}\ y_{i}$ is at most $b_j\ (d-1)\ a_{\ell+1}$.
Hence, the increasing rate of the dual is at most:
\begin{align*}
& b_j\ (d-1)\ a_{\ell+1} + b_j \left(1 - f_d\biggl( \frac{\ell}{d} \biggr) \right) \\
&= b_j\ (d-1)\ a_{\ell+1} + b_{j} \left(1 - f_d\biggl( \frac{\ell+1}{d} \biggr) + a_{\ell+1} \right) \\
&= b_j\ d\ a_{\ell+1} + b_{j} \left(1 - f_d\biggl( \frac{\ell+1}{d} \biggr) \right)
= \frac{b_j}{C(d)}
\end{align*}
where the first equality holds since $f_d\bigl( \frac{\ell+1}{d} \bigr) = f_d\bigl( \frac{\ell}{d} \bigr) + a_{\ell+1}$
and the last equality follows \Cref{eq:allocation-case1}.

The increasing rate of the dual is at most $b_{j}/C(d)$, while for the primal it is $b_{j}$. The lemma follows from this conclusion.
\end{proof}

\end{document}